\documentstyle[amssymb,12pt,psfig]{article}

 \newcommand{\be}{\begin{equation}}
  
 \newcommand{\ee}{\end{equation}}
  
 \newcommand{\ba}{\begin{eqnarray}}
  
 \newcommand{\ea}{\end{eqnarray}}

\begin{document}

\title{Bose-Einstein condensation in real space}
\author{J.J. Valencia, M. de Llano \\
Instituto de Investigaciones en Materiales, UNAM \\
Apdo. Postal 70-360, 04510 M\'{e}xico, DF, Mexico\\
M.A. Sol\'{\i}s \\
Department of Physics, Washington University \\
St. Louis, Missouri 63130, USA and \\
Instituto de F\'{\i}sica, UNAM, Apdo. Postal 20-364,\\
01000 M\'{e}xico, DF, Mexico}

 \maketitle

 \begin{abstract}
 We illustrate how Bose-Einstein condensation occurs not only in momentum
 space but also in coordinate (or real) space. Analogies  between the
 isotherms of a van der Waals gas of extended (or finite-diameter) identical
 atoms and the point (or zero-diameter) particles of an ideal Bose gas allow
 one to conclude that, in contrast to the van der Waals case, the volume per
 particle can go to zero in the pure Bose condensate phase precisely because
 the particle diameter is zero.
 \end{abstract}

 PACS \# 03.75.Hh; 05.30.Jp; 05.70.Fh

 \pagebreak

 \section{Introduction}

 It is sometimes said that the Bose-Einstein condensation (BEC) in a perfect
 or ideal (i.e., without interactions) boson gas is a condensation in
 momentum space, but not in coordinate or real space like the condensation of
 vapor into liquid. For example: \noindent F. London \cite{Londonphysrev}
 claims that {\it ``...one may say that there is actually a condensation, but
 only in momentum space, and not in ordinary space, ...[where] no separation
 of phases is to be noticed.''} \cite{Londonsuperfluids}. The same author
 speaks of bosons that {\ ``...}{\it settle in some kind of order in momentum
 space even at the expense of order in ordinary space.'' } \noindent Landau
 \& Lifshitz \cite{LyL} state that: {\it ``The effect of concentrating the
 particles in the state $\epsilon =0$ is often called `BEC'. We must
 emphasize that at best one might perhaps talk about `condensation in
 momentum space.' Actual condensation certainly does not take place in the
 gas.''} \noindent T.L. Hill \cite{Hill} asserts that {\it ``...As it is
 usually stated, the condensation occurs in momentum space rather than in
 coordinate space: the condensed phase consists of molecules with zero energy
 and momentum, and macroscopic de Broglie wavelength.''} \noindent Fetter \&
 Walecka \cite{FyW} say this: {\it ``...The assembly is ordered in momentum
 space and not in coordinate space; this phenomenon is called BEC.''}
 \noindent B. Maraviglia \cite{Maraviglia} writes (freely translated) that 
 {\it ``...Superfluity results from the fact that the }${\it ^{4}He}${\it \
 atoms, since they obey BE statistics, can condense not in position but in
 momentum space....''} \noindent F. Mandl \cite{Mandl} says {\it ``...It
 differs from the condensation of a vapour into a liquid in that no spatial
 separation into phases with different properties occurs in BEC.''} Finally,
 D.A. McQuarrie {\cite{McQ} p. 176} concludes {\it ``...Therefore the BEC is
 a first-order process. This is a very unusual first-order transition,
 however, since the condensed phase has no volume, and the system therefore
 has a uniform macroscopic density rather than the two different densities
 that are usually associated with first-order phase transitions. This is
 often interpreted by saying that the condensation occurs in momentum space
 rather than coordinate space,....''}

 We argue here that Bose-Einstein condensation is a phase transition that
 occurs in real space too, which substantiates assertions made by other
 authors, e.g., R. Becker \cite{BeckerZ}, D. ter Haar \cite{tH}, K. Huang %
 \cite{Huang} and D.L. Goodstein \cite{Goodstein}.

 \section{Van der Waals gas}

 The van der Waals equation of state for a clasical monatomic gas is 
 \begin{equation}
 \left[ P+a\left( \frac{N}{V}\right) ^{2}\right] (V-Nb)=Nk_{B}T,  \label{ev}
 \end{equation}%
 where $P$ is the pressure, $V$ the volume, $T$ the absolute temperature, $N$
 the number of atoms and $k_{B}$ Boltzmann's constant. The effective
 ``excluded volume'' per particle \cite{Lee} is 
 \begin{equation}
 b=\frac{1}{2}\frac{4}{3}\pi (\sigma )^{3}=\frac{2}{3}\pi \sigma ^{3};
 \label{bp}
 \end{equation}%
 where $\sigma $ is the diameter of each particle, thought of as a hard
 sphere. It is the reduction in the original volume per particle $V/N$ due to
 finite-sized atoms, and was proposed by Clausius for an imperfect gas \cite%
 {Lee}. In 1873 van der Waals introduced a second correction term (see e.g.,
 Ref. \cite{Lee}) to the equation of state $PV=Nk_{B}T$ of an ideal gas to
 account for the attractive forces between molecules. In (\ref{ev}) the
 parameter $a$ is given by 
 \begin{equation}
 a\equiv -\frac{4\pi }{2}\int_{\sigma }^{\infty }u(r)r^{2}dr,  \label{a}
 \end{equation}%
 where $u(r)\leq 0$ is the attractive interaction potential between two atoms
 whose center-to-center separation is $r$.

 On a $P-V$ phase diagram (\ref{ev}) exhibits the well-known isotherm loops
 signaling a vapor to liquid phase transition. One such loop (at a given $T$)
 is shown in Fig. 1 (left panel), where the horizontal plateau connecting
 points D and B is called the ``Maxwell construction.'' Loops occur only for
 isotherms with $T<T_{c}$, where $T_{c}$ is the critical point where both $%
 (dP/dV)_{T}=0$ (zero slope) and $(d^{2}P/dV^{2})_{T}=0$ (change of
 curvature). These two conditions along with (\ref{ev}) give 
 \begin{equation}
 P_{c}=\frac{a}{27b^{2}};\hspace{2cm}V_{c}=3Nb;\hspace{2cm}T_{c}=\frac{8a}{%
 27k_{B}b}  \label{vc}
 \end{equation}%
 for the critical pressure, volume and temperature. 

\begin{figure}[bh]
\begin{minipage}[b]{3.0in}
\psfig{file=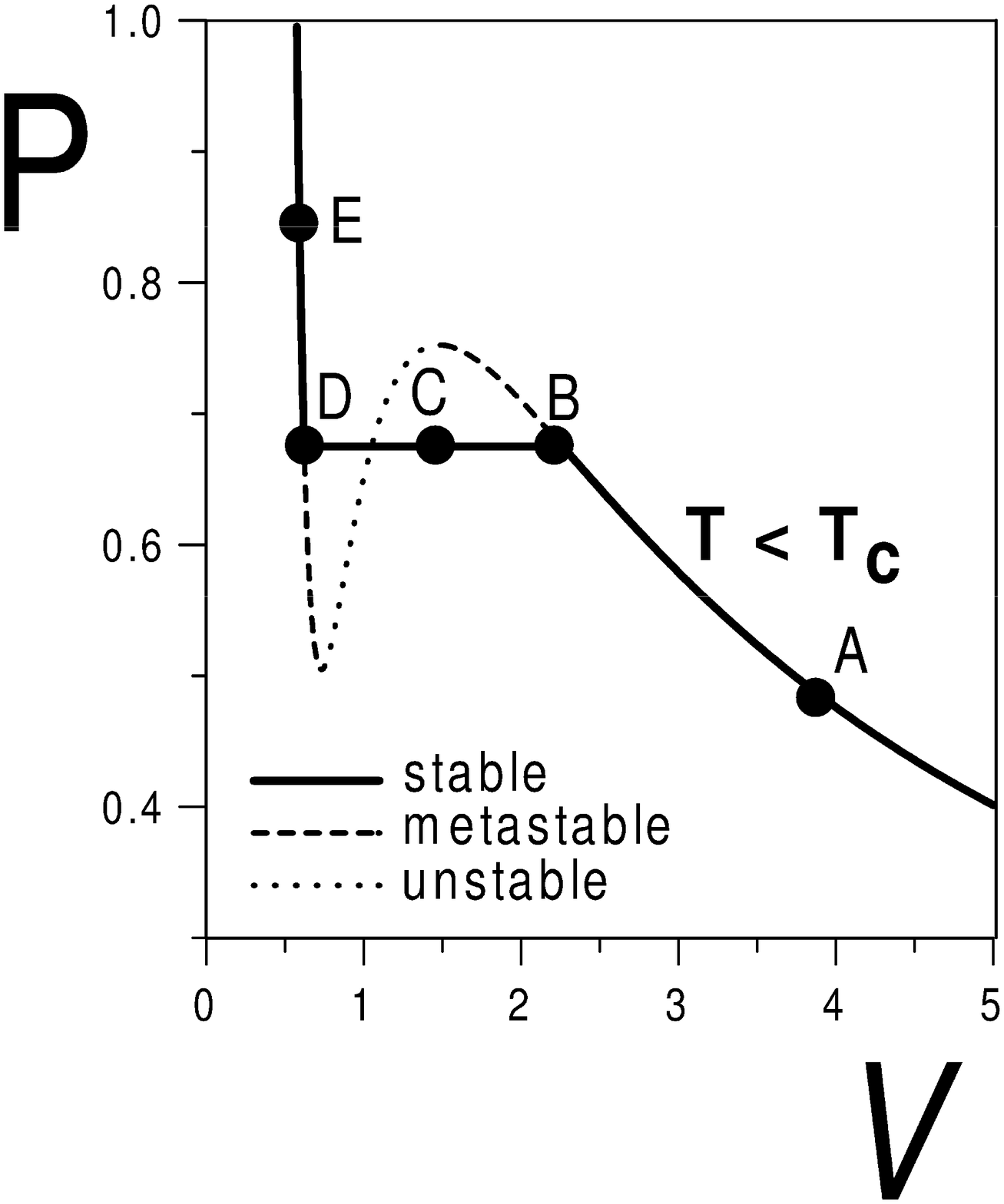,height=3.20in,width=3.0in}
\end{minipage} \hfill 
\begin{minipage}[b]{3.50in}
\psfig{file=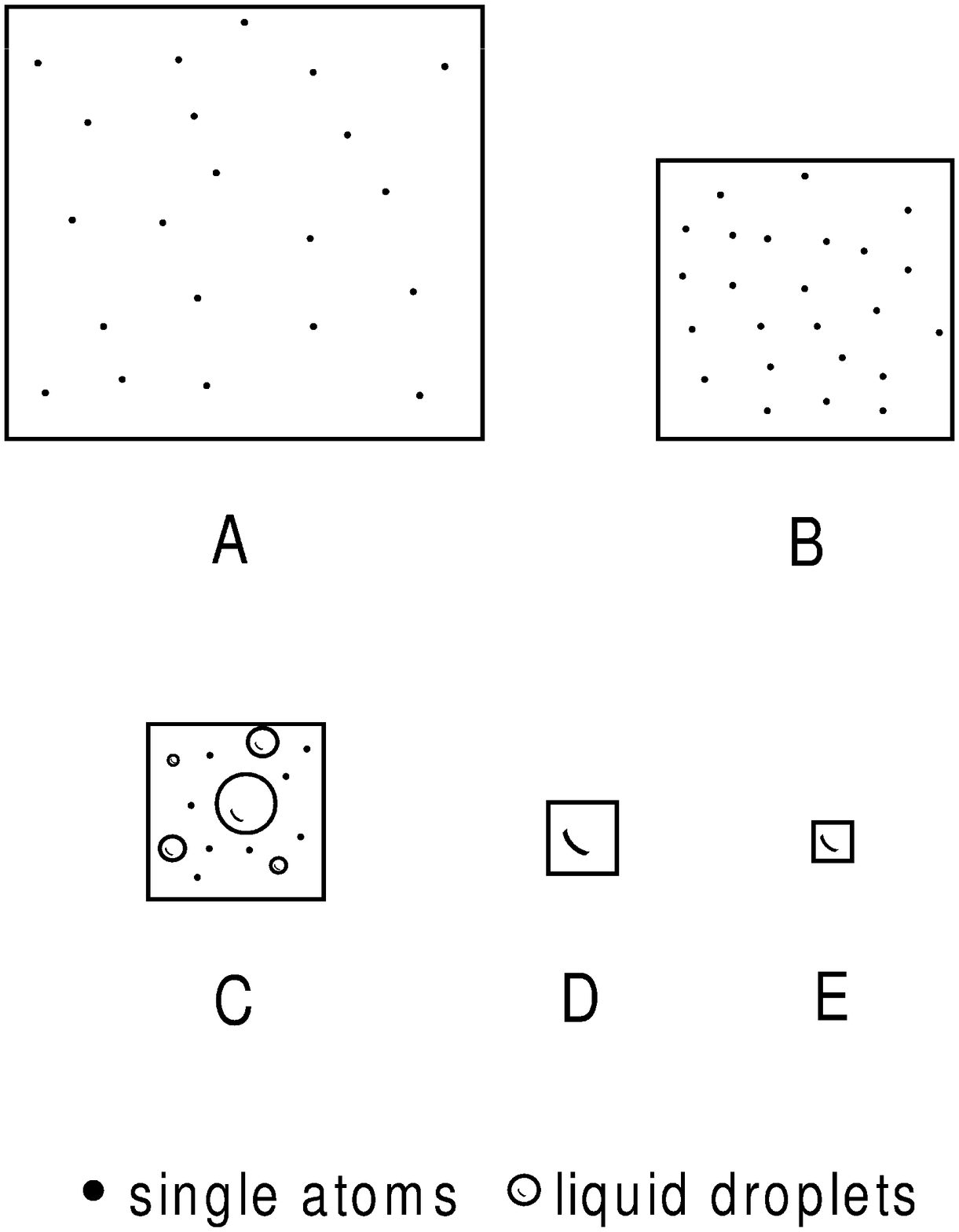,height=3.50in,width=3.0in}
\end{minipage}
\begin{quotation}
\caption{ {\bf Left}: Schematic sketch of a typical van der
Waals isotherm below the critical tempetature $T_c$ in the pressure-volume
plane (in arbitrary units) for a classical monatomic fluid of finite-sized
atoms. The horizontal plateau DCB corresponds to Maxwell's construction,
which separates ``stable'' from ``metastable'' states, the latter being
separated by an ``unstable'' portion as shown. {\bf Right}: illustration of
how, as the volume of the system is reduced from points A to B to C and D
along the chosen isotherm on the left, the vapor condenses first into
several ``droplets'' of different sizes and finally into a single
``self-bound'' drop, D and E. All points such as E correspond to a sharp
rise in pressure because the single drop at D is being compressed as volume
is reduced.}
\end{quotation}
\end{figure}

 \subsection{BEC as a condensation in real space}

 For a quantum ideal gas in three dimensions 
 \begin{equation}
 PV=\frac{2}{3}U,  \label{pv}
 \end{equation}%
 where $U$ is the internal energy, if \cite{Navarro}\ a quadratic
 energy-momentum (or dispersion) relation holds for each particle. If $T_{c}$
 is the BEC transition temperature below which there is macroscopic
 occupation in a given single quantum state (not mix up with critical temperature for a van der Waals gas), the internal energy for the
 ideal Bose gas for $T\leq T_{c}$ (or alternatively $V\leq V_{c}$ where $V$
 is the bosonic system volume and $V_{c}$ the transition volume) is given in
 Ref. \cite{Huang}, Eq. (12.62). If we substitute Eqs. (12.55) into (12.62)
 of Ref. \cite{Huang} we obtain 
 \begin{equation}
 \frac{U(V,T)}{Nk_{B}T}=\frac{3}{2}\frac{g_{5/2}(1)}{g_{3/2}(1)}\left( \frac{T%
 }{T_{c}}\right) ^{3/2}\,\,\,\,\,\,\hbox{for all}\,\,V\leq V_{c}.
 \label{3.20d=3s=2}
 \end{equation}%
 Here $g_{\sigma }(z)$ are the so-called Bose functions defined (see p. 506
 of Ref. \cite{Path}) as 
 \begin{equation}
 g_{\sigma }(z)\equiv \frac{1}{\Gamma (\nu )}\int_{0}^{\infty }dx\frac{x^{\nu
 -1}}{z^{-1}e^{x}-1}=\sum_{l=1}^{\infty }\frac{z^{l}}{l^{\sigma }},
 \label{3.PD2}
 \end{equation}%
 where the ``fugacity'' $z\equiv e^{\mu /k_{B}T}$, with $\mu $ the boson
 chemical potential. It is well-known that the series 
 \[
 g_{\sigma }(1)=\sum_{l=1}^{\infty }\frac{1}{l^{\sigma }}\longrightarrow
 \infty \,\,\hbox{when}\,\,\sigma \leq 1,
 \]%
 since for $\sigma =1$ we have $g_{1}(1)=1+\frac{1}{2}+\frac{1}{3}+...$, the
 familiar harmonic series which diverges. On the other hand, for $\sigma >1$ 
 \begin{equation}
 g_{\sigma }(1)=\sum_{l=1}^{\infty }\frac{1}{l^{\sigma }}\equiv \zeta (\sigma
 ),\hspace{2cm}(\sigma >1),  \label{zr}
 \end{equation}%
 the Riemann-Zeta function. Thus, (\ref{pv}) and (\ref{3.20d=3s=2}) give 
 \begin{equation}
 P=\frac{2}{3}\frac{U}{V}=\frac{\zeta (5/2)}{\zeta (3/2)}\frac{Nk_{B}T}{V}%
 \left( \frac{T}{T_{c}}\right) ^{3/2}\hspace{0.3cm}(\hbox{for}\hspace{0.3cm}%
 V\leq V_{c}).  \label{P}
 \end{equation}%
 If we use for the thermal wavelength $\Lambda \equiv h/\sqrt{2\pi mk_{B}T}$
 and Eq. (10.58) of Ref. \cite{McQ}, the condensate fraction for $T\leq T_{c}$
 is 
\begin{equation}
\frac{N_{0}(T)}{N}=1-\frac{\zeta (3/2)}{8\pi ^{3/2}(\hbar
^{2}/2m_{B}k_{B}T)^{3/2}(N/V)},  \label{3.10a}
\end{equation}%
 where $N_{0}(T)$ is the condensate particle number and $N$ the total
 particle number.

 Using the fact that $N_{0}(T)$ is negligible compared with $N$ when $ T\geq
 T_{c}$, (\ref{3.10a}) leads to the well-known BEC $T_{c}$ formula 
 \begin{equation}  \label{3.2.3}
 T_c = \frac{\hbar^2}{2mk_B} \left[\frac{8 \pi^{3/2} N/V} {\zeta(3/2)}\right]%
 ^{2/3}\simeq 3.313\frac{\hbar ^{2}}{m_{B}k_{B}}\left(\frac{N}{V}%
 \right)^{2/3},
 \end{equation}
 since $\zeta (3/2)\simeq 2.612$. Alternatively, from (\ref{3.10a}) the
 critical volume $V_c$ below which the BEC appears at any temperature $T$ is 
 \begin{equation}
V_c=\frac{(\hbar^2/2m_Bk_BT)^{3/2}8\pi ^{3/2}N}{\zeta(3/2)}.  \label{vt}
\end{equation}
 Combining (\ref{3.2.3}) with (\ref{P}) leaves 
 \begin{equation}
 P=\frac{2}{3}\frac{U}{V}=\frac{\zeta (5/2)}{\sqrt{(2\pi)^3}}\left(\frac{%
 \sqrt{m_B}}{\hbar}\right)^3 (k_BT)^{5/2}\simeq 0.0851\left(\frac{\sqrt{m_B}}{%
 \hbar}\right)^3(k_{B}T)^{5/2},\,\,\,\,\, \,\hbox{for all}\,\,V\leq V_{c},
 \label{AB. P}
 \end{equation}
 since $\zeta (5/2)\simeq 1.341$. So, at  constant temperature $T$ if we
 reduce the volume below the value $V_c$ given by (\ref{vt}), the pressure
 stays constant. This corresponds to the portion BCD of the isotherm depicted
 in the left panel of Fig. 2. The condensate fraction given by (\ref{3.10a}),
 combined with (\ref{vt}), simplifies to 
 \begin{equation}
 \frac{N_{0}(T)}{N}=1-V/V_c\hspace{1cm} \hbox{for all} \hspace{.5cm}V\leq V_c.
 \label{3.13}
 \end{equation}

 \section{Zero volume a sign of real-space condensation}

 Imagine the ideal Bose gas to be in a cylinder with a movable piston. 
 According to (\ref{AB. P}), if we push the piston in, decreasing the 
 available volume below $V_c$, given by (\ref{vt}), at constant temperature $T
 $, the pressure remains constant. The piston can be pushed in at constant
 pressure until the two-phase region BCD of Fig. 2 vanishes, i.e., until the
 condensate particle number $N_0$ equals the total number of particles $N$.
 Thus, at B the condensate just begins to appear and at D there is $100\%$
 condensate.

 So, to have BEC in coordinate space the gas must be condensed in momentum
 space, i.e., a macroscopic number of bosons must be in the ground state. The
 whole gas occupies zero volume only if the bosons are not moving with
 different speeds and directions. In the two-phase region, where $N\neq
 N_{0}\neq 0$ the condensed phase consisting of several zero-diameter
 ``droplets'' with different particle numbers do not occupy any volume at
 all. Fig. 2 shows how the volume becomes zero when the bosonic system is
 entirely condensed at point D; see also (\ref{3.13}) when $N=N_0$. However,
 for a van der Waals fluid when the vapor is entirely condensed into liquid
 at D, Fig. 1, the volume cannot be zero because of finite particle sizes,
 and the pressure rises steeply to points E and beyond as the particles are
 further compressed against each other.

 We have apparently fallen into a contradiction since $N = N_0$ usually
 applies to an ideal bosonic system at $T = 0$, and not to a system along a
 finite $T$ isotherm. However, we note that in keeping with $N \to N_0$ for
 some $T \neq 0$, we approach the endpoint of the two-phase region  where 
 {\it all} the isotherms merge together, in particular the $T_c$ isotherm
 with which we began as well as the $T = 0$ isotherm.

\begin{figure}[bh]
\begin{minipage}[b]{3.50in}
\psfig{file=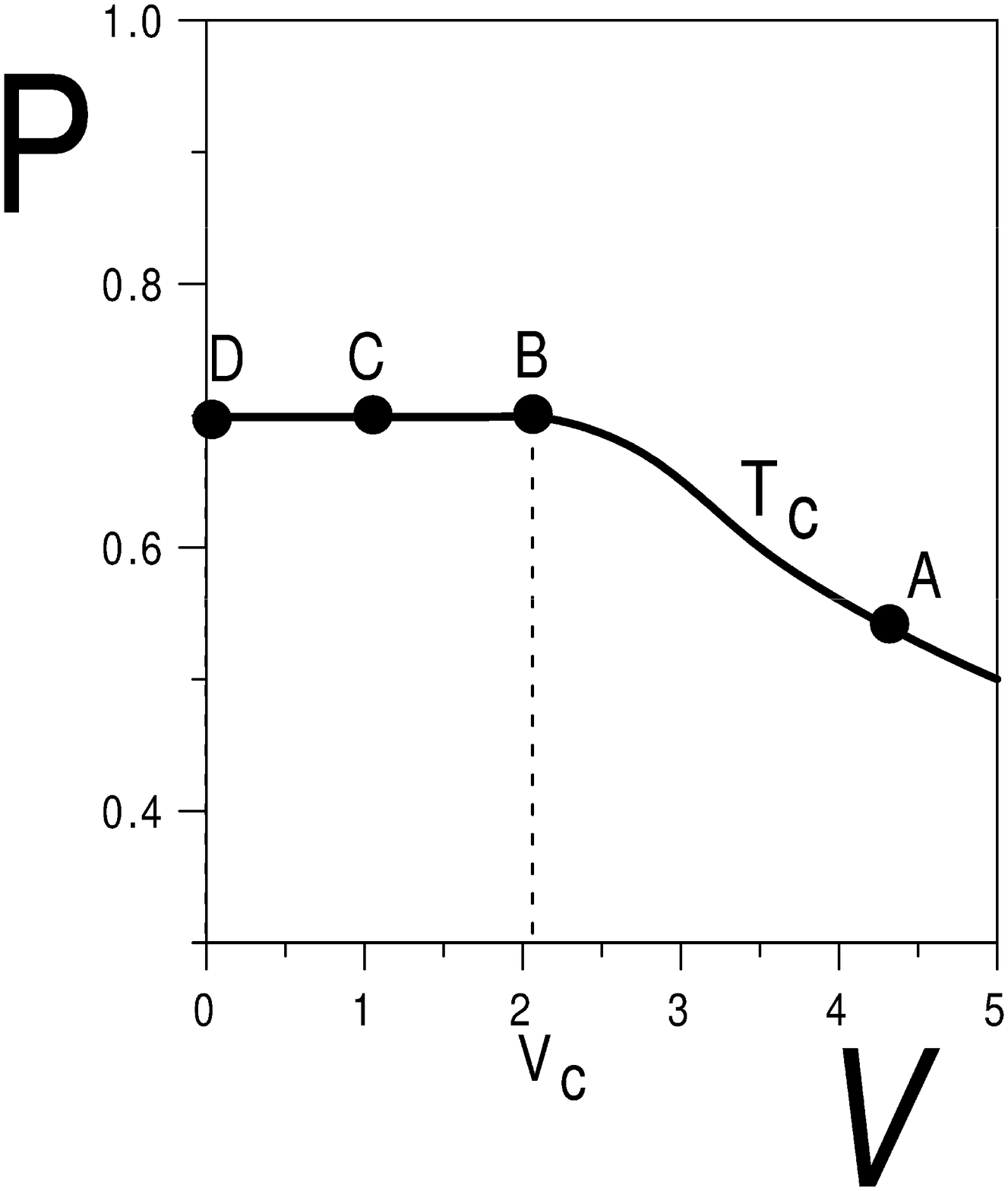,height=3.5in,width=3.00in}
\end{minipage} \hfill
\begin{minipage}[b]{3.50in}
\psfig{file=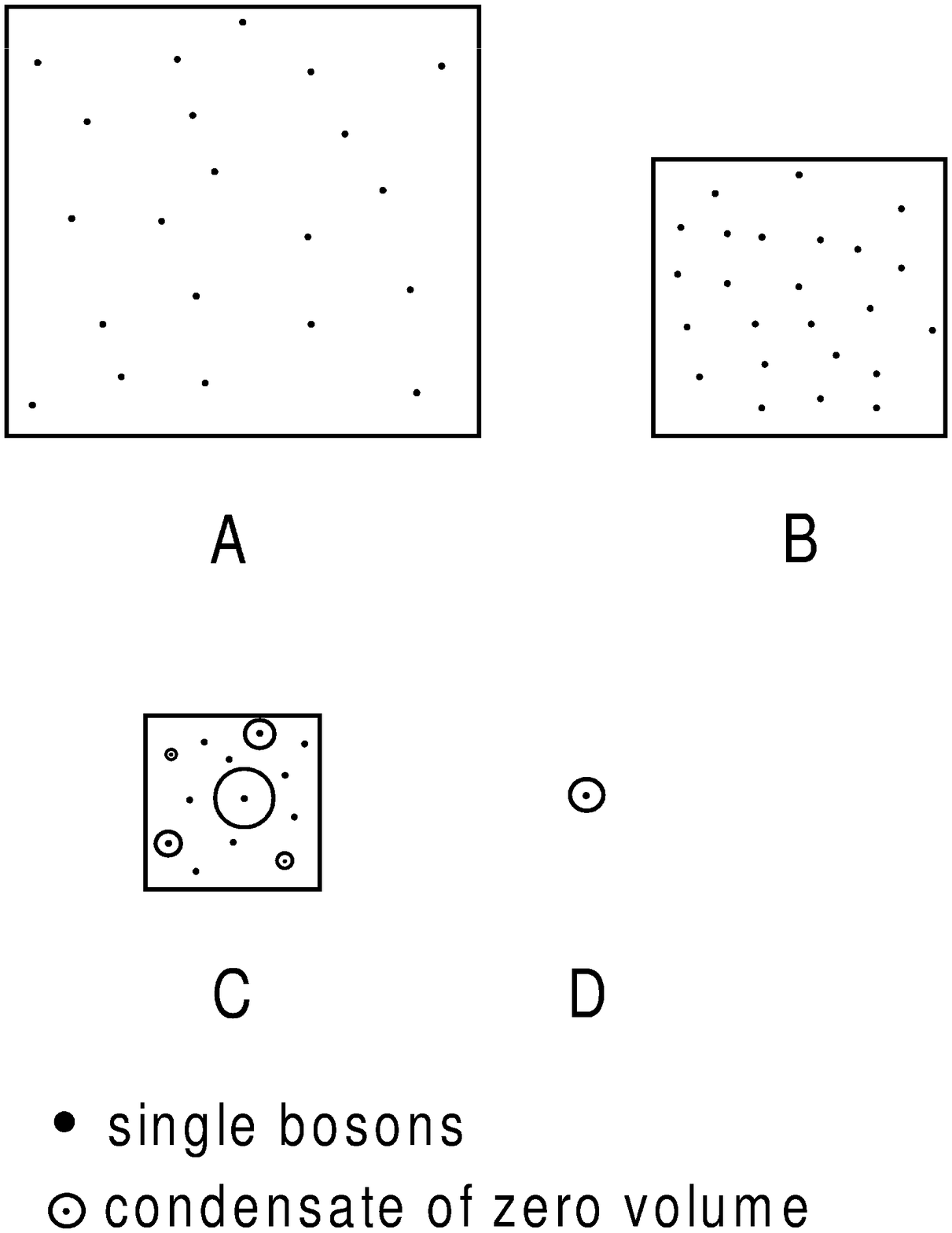,height=3.60in,width=3.0in}
\end{minipage}
\begin{quotation}
\caption{ {\bf Left}: schematic isotherm in the $P-V$ plane
(in arbitrary units) for an ideal Bose gas at some fixed $T=T_c$ as given by
(\ref{3.2.3}). Being ideal, the gas consists of {\it zero}-diameter
particles, i.e. with zero-range interparticle repultions. {\bf Right}:
illustration of how system behaves at different volumes marked as A, B, C
and D on the isotherm, with circled dots of varying sizes denoting different
sized condensates of zero volume (since the bosons are point particles).}
\end{quotation}
\end{figure}

 \section{Conclusions}

 By analogy with a van der Waals gas of zero-diameter atoms we have
 illustrated how Bose-Einstein Condensation (BEC) occurs not only in momentum
 space, i.e., $N=N_0$, but also in real space if we applied a external potential. This vindicates the following
 claims:

 \begin{itemize}
 \item R. Becker \cite{BeckerZ} p. 120, {freely translated: {\it ``...the
 number of atoms in the condensate phase, $N_{0}$, exhibits null
 volume....''}}

 \item D. ter Haar \cite{tH} ``...{\it one can also consider Einstein 
 condensation to be a condensation in coordinate space.''}

 \item K. Huang \cite{Huang} p. 290 ``...{\it If we examine the equation of
 state alone, we discern no difference between the BEC and an ordinary
 gas-liquid condensation}....''

 \item D.L. Goodstein \cite{Goodstein} p. 132 ``...{\it condensation takes
 place in real as well as momentum space}....''
 \end{itemize}

\end{document}